\documentclass[aps,prl,twocolumn,showpacs,superscriptaddress,amsmath,amssymb,nofootinbib]{revtex4}

\usepackage{graphicx}

\begin{document}



\title{Are There Many Worlds?}



\author{Austin Gerig} \email{austingerig@gmail.com}
\affiliation{CABDyN Complexity Centre, University of Oxford, Oxford OX1 1HP, United Kingdom}


\date{September 27, 2014}

\begin{abstract}

It is often thought that the existence of other worlds cannot be scientifically verified and therefore should be treated as philosophical speculation.  In this article, I describe several methods for determining if other worlds exist, even without interacting with them.  These methods are based on the following premise: if there are many worlds, then the statistical properties of a natural process are biased when measured by an observer whose existence was influenced by the process.  The bias is always in the same direction, making the process appear more beneficial for the existence of the observer than it actually is.  I suggest several potential ways of measuring the bias, show through a simple model of population dynamics how the bias is generated, and briefly consider whether our current drop in population growth is evidence of many worlds.

\end{abstract}



\maketitle



\section{Introduction}

Several well-known theories in physics posit that the universe is extremely large, containing a vast number of causally disconnected regions or ``worlds''\cite{Everett1957,DeWitt1973,Guth1997,Garriga2001,Tegmark2003,Vilenkin2006}. Suppose these theories are correct and there are many worlds. How could we possibly know they exist?  In this article, I show that if there are many worlds, then the environment of a typical observer will be biased -- it will appear more beneficial for their existence than it should be.  Assuming we are typical observers, the presence of such bias in our own environment is evidence of many worlds.   

The results I present are based on a modified and more complete version of the anthropic principle:
\begin{quote}
When interpreting data, you must account for the fact that your location in the universe is necessarily privileged: First, you will only find yourself in a location where it is possible that you exist.  Second, you are more likely to find yourself in a location where it is more probable that you exist.
\end{quote}
Both statements are truisms.

The first truism is simply a rephrasing of Carter's anthropic principle\cite{Carter1974,Carter1983,Dicke1961}.  It serves as a useful reminder that your surroundings have been subject to a selection effect: out of the entire universe, your location has been ``selected'' to be compatible with your existence.  In practice, this means you should not simply assume you live in a typical region of the universe.  It could be that you live in an atypical or ``surprising'' region because a typical region is incompatible with you.  

The second truism is new in the sense that it has never been formally stated.  It accounts for the second form of ``privilege'' endowed to your location -- you are more likely to find yourself in a location where it is more probable that you exist.  It serves as a useful reminder that if you believe certain locations are more conducive for your existence than others, you must account for this when interpreting data.  In practice, this means that \emph{even among those regions of the universe that are compatible with your existence}, you should not assume your region is typical.  It could be that some compatible regions are more conducive for your existence (i.e., are ``biased'' towards you), and you are more likely to find yourself in one of these regions simply because they are more likely to produce you.

The truisms are just that -- they are true by construction.  They do not attempt to formalize how to determine which locations in the universe are compatible with your existence nor how you should determine the likelihood of your existence across these locations.  Instead, they are important reminders that \emph{if} you believe certain regions are more likely to produce you than others, then you \emph{must} account for this when interpreting data.\footnote{Eddington was perhaps the first physicist to make use of the complete version of the anthropic principle when he stated, ``It is practically certain that a universe containing mathematical physicists will at any assigned date be in the state of maximum disorganization which is not inconsistent with the existence of such creatures.''\cite{Eddington1931} He was using anthropic reasoning to point out a deep problem with Boltzmann's hypothesis that our region of the universe is a thermal fluctuation -- if true, then we are much more likely to exist within a small and brief fluctuation than within a large fluctuation the size of our world.}

In this article, I am interested in using anthropic reasoning to determine if many worlds exist.  To derive results, I will add an assumption to the second truism that is accepted by nearly all physicists and philosophers who have written on the topic\cite{Eddington1931, Barrow1986, Vilenkin1995, Leslie1990, Olum2000, Bostrom2002, Bostrom2003, Albrecht2004, Tegmark2005, Neal2006, Page2007, Weinberg2007, Pogosian2007, Page2008, Garriga2008, Bousso2008, Cirkovic2010, Waltham2011, Gerig2012, Gerig2013}, but that \emph{could be} incorrect (see \cite{Hartle2007, Srednicki2010}) -- I will assume that you are more likely to exist at locations in the universe where many observers like you have been generated.  Specifically, I will assume that the likelihood of your existence at one location vs.~another is in direct proportion to the ratio of suitable observers existing at those locations.

There are several ways to motivate this assumption. One such way is through an appeal to typicality across a suitable reference class of observers, or to what has been called the principle of mediocrity\cite{Vilenkin1995, Garriga2008}.  Without evidence to the contrary, you should assume you are typical within each reference class to which you belong -- as if you were randomly selected from all individuals within that reference class.\footnote{A similar assumption, called the Self-Sampling Assumption (SSA), exists in philosophy and is motivated using thought experiments\cite{Bostrom2002}.  There is also the Self-Indication Assumption (SIA) in philosophy, which extends a reference class from existing individuals within that class to all individuals within the class, whether they exist or not\cite{Bostrom2002}.  Under either the SSA or SIA, the assumption I make holds true -- that the likelihood of your existence at one location vs. another is in direct proportion to the ratio of suitable observers existing at those locations.}  If you are a random selection across suitable observers, then the likelihood that you exist at one location vs. another is in direct proportion to the ratio of suitable observers at those locations. 

Another, perhaps more natural, way to motivate the assumption is to treat your existence as the result of an experiment by the universe.  At various locations, the universe has generated intelligent observers, each of which had a certain prior potential of being you.  If we assume each generated observer had equal potential of being you, then the likelihood that you exist at one location vs. another is in direct proportion to the ratio of observers at those locations.\footnote{Note that a particularly useful consequence of this reasoning is that you can assume different potentials across observers (for example, you may think that an alien or a monkey is much less likely to have the necessary brain structure required to be you).  In this case, you can appropriately do anthropic reasoning by weighting observers according to these different potentials, which is something that cannot be done when using reference classes (see footnote 6 in \cite{Gerig2012}).}

Throughout the article, I will use the term ``the universe'' to refer to the set of all things that have existed or that will ever exist (including all of space-time), the term ``small universe'' to represent that the universe is limited in size and age so that it does not extend far beyond its current observable edge and will not last forever, and the term ``large universe'' to represent that the universe is extremely large (perhaps infinitely large) and contains many other disconnected worlds, either via infinite space-time, separate inflationary events, quantum branching, and/or some other mechanism. 

As I show below, if the universe is large, then you will typically find yourself at a location that is atypically suited for you, i.e., at a location that appears unusually biased towards your existence.  This bias is all-encompassing, influencing all processes that affect the probability of your existence.  For example, you are likely to observe a higher than baseline rate of population growth in the past, a lower than baseline rate of catastrophic risk\cite{Tegmark2005,Cirkovic2010}, physical constants that are not only compatible but especially conducive for galaxy formation\cite{Weinberg1987,Vilenkin1995,Weinberg2007,Pogosian2007}, and an evolutionary history that appears unnaturally guided to your existence.

The anthropic principle has been used in other papers to argue for a ``multiverse''\cite{Leslie1988,vanInwagen1993,Bradley2009}. These papers note that our location in the universe seems fine-tuned to our existence, i.e., out of the set of possible worlds with our physical laws, only a miniscule fraction have physical constants that are compatible with the development of life\cite{Carr1979,Barrow1986,Hogan2000}.  That we exist, therefore, is evidence that the universe has sampled from the space of possible worlds many times, and therefore must be quite large.  I will discuss this argument and develop a more general version of it below.

The rest of the article is organized as follows.  I first consider whether your existence itself is evidence of a large universe.  I then show that if the universe is large, you will typically exist at a different location than if the universe is small.  Finally, as a proof of concept, I model population dynamics and show that a typical observer in a universe of many worlds will observe a strong bias in population growth that is removed after they exist, but a typical observer in a universe of one world will not.

\section{Should you exist?}

The simplest argument for a large universe is that it makes your existence certain, whereas otherwise it might be an extremely surprising result.  Suppose that you are unlikely to exist if the universe is small but that you are certain to exist if the universe is large.  Your current existence, therefore, is evidence that the universe is large.  

More formally, assume $P(E|S) \ll 1$ and $P(E|L) = 1$, where $E$ denotes that you exist, $S$ denotes that the universe is small, and $L$ denotes that the universe is large.  Assuming your prior for a large universe is not infinitesimal, then after considering that you exist, the posterior odds of a large universe are,
\begin{equation}
\frac{P(L|E)}{P(S|E)} = \frac{P(E|L)P(L)}{P(E|S)(1-P(L))} \gg 1,
\end{equation}
and you should believe with almost certainty that the universe is large.  

The above argument depends critically on whether or not your existence is rare in a small universe.  It also requires that you treat your existence as a pre-determined possible outcome, i.e., as if you were a specific lottery ticket that may or may not be drawn by the universe.

Notice, however, that the argument is very general.  It works regardless of the reason \emph{why} your existence is rare.  For example, and as previously mentioned, you may think that a small universe is extremely unlikely to contain a world with the correct physical constants necessary for your existence.  Alternatively, you may think that in a small universe, life cannot evolve under the time constraints imposed by stellar life cycles.     

Even if you believe it likely that people exist in a small universe, you can still be convinced that the universe is large if you believe it unlikely that \emph{you} were produced by the universe.  For example, you might think that the number of possible people is astronomically large and that a small universe surely could not sample from this space enough times to make it likely that \emph{you} exist.  Such an argument requires only the assumption that the space of distinct possible people is very large, and it works even if the universe was designed specifically for intelligent life. 

\section{Where should you be located?}
Suppose you think your existence is certain, even in a small universe, so that the previous argument is not compelling.  In this case, there are other ways to find evidence that the universe is large.  Here, I assume you exist and show if the universe is large, you will typically exist in a different location than if the universe is small.  Therefore, in principle, you should be able to determine whether or not the universe is large by carefully observing your surroundings.

Consider the following toy cosmological model with two types of worlds\cite{Page1999, Page1999b, Page2011},
\begin{itemize}
\centering
\item[] World $W_1$: 1 million observers, $\mathcal{P}(W_1) = 0.999$\\
World $W_2$: 1 trillion observers, $\mathcal{P}(W_2) = 0.001$
\end{itemize}
If the universe is small and draws only one world, then with 99.9\% probability it will draw a world of type $W_1$, and you will exist there.  However, if the universe is large and draws both types of worlds many times, then you are much more likely to exist in $W_2$ than in $W_1$.  The $W_2$ worlds comprise only a tiny fraction of a large universe, but they are much more likely to produce you: they sample from the space of observers 1 million times more frequently than $W_1$.

More formally, assume the set of all physically possible worlds, $\Omega$, is completely specified.  Furthermore assume the laws of physics are known and they determine the probability measure $\mathcal{P}$ on $\Omega$ which represents the outcome of nature selecting one of these worlds for existence.  The universe is the collection of worlds that exist, i.e., the set of possible worlds selected by nature.  Denote by $\Omega'\subseteq \Omega$ the subset of possible worlds where you exist, and assume it is certain you exist, i.e., $\sum_{\omega\in\Omega'} \mathcal{P}(\omega) = 1$.  Finally, denote by $n(\omega)$ the number of observers that exist in possible world $\omega$.

In a small universe, the world in which you live is simply chosen according to the laws of nature.  The probability that you find yourself in world $\omega$ is therefore,
\begin{equation}
P_S(\omega) = \mathcal{P}(\omega).\label{eq.small}
\end{equation}

In a large universe, many worlds exist.  Assuming a large universe has sampled many times from the set of possible worlds and that you are a typical observer across these worlds\cite{Page2007, Garriga2008}, then the probability that you find yourself in world $\omega$ is,
\begin{equation}
P_L(\omega) = \frac{n(\omega)\mathcal{P}(\omega)}{\sum_{\omega\in\Omega'}{n(\omega)\mathcal{P}(\omega)}}.\label{eq.large}
\end{equation}
Notice that in general, $P_S(\omega)\neq P_L(\omega)$, and you typically will find yourself in a different world depending on whether or not the universe is large.  

In a large universe, your location is biased towards worlds that are conducive for your existence.  The bias can be subtle.  For example, if a certain medical discovery allows your civilization to produce twice as many observers, then you are twice as likely to exist in a world where your civilization makes the discovery than in one where it doesn't.  The bias can also be quite strong.  It can be strong enough such that you exist in a world that is highly atypical -- a world that you would not find yourself in if the universe only consisted of one world.  

Going back the original example and using Eqs.~\ref{eq.small} and \ref{eq.large},
\begin{eqnarray}
P_S(\omega\in W_1) & = & 0.999, \\
P_S(\omega\in W_2) & = & 0.001, \\
P_L(\omega\in W_1) & \approx & 0.001, \\
P_L(\omega\in W_2) & \approx & 0.999.
\end{eqnarray}
Notice that the probability of being in one or the other type of world is reversed if the universe is small vs large, and that your world type, therefore, can be used as convincing evidence that the universe is small or large.  

As previously mentioned, the above calculations assume your existence is certain whether or not the universe is small or large.  If it is unlikely you exist in a small universe (perhaps because it is unlikely you exist if only a small number of observers typically exist), then the analysis leading to Eq.~\ref{eq.small} would be different (see \cite{Olum2000} and \cite{Gerig2012}).  However, if it is unlikely you exist in a small universe, then your existence by itself should convince you that the universe is large (see Section II), and you should use Eq.~\ref{eq.large} instead.

\section{Determining Bias}

Although it is relatively straightforward to demonstrate that a typical observer in a large universe will exist in a world biased towards their existence, it is perhaps not as easy for an observer to determine that the bias exists.  The observer would need a correct model of cosmology, whereas often cosmological models are formulated by extrapolating from an observer's surroundings.  For example, in the above toy model, an observer in a large universe who exists in $W_2$ might interpret their circumstance as evidence that a $W_2$ world is typical rather than as evidence that the universe is large.

There are several ways to overcome this problem and establish that your world is biased.  I describe four methods below.  The first two methods search for evidence of bias in processes that are inherently unobservable and that you must posit exist, whereas the second two attempt to measure bias in processes that are observable or that you can collect direct evidence of.

\textbf{(1)} As previously mentioned, you might discover that certain parameters in your world are finely-tuned to values that are compatible with your existence.  Assuming these parameters could have taken a number of different values, finding that they are set within a narrow range that allows your existence is compelling evidence of bias.  For example, physical constants and biological parameters appear finely-tuned to our existence\cite{Carr1979,Barrow1986,Hogan2000,Carter1983} 

\textbf{(2)} In addition to being compatible with your existence, you might find that parameters in your world are tuned even further, to values that are especially conducive for your existence.  Again, assuming these parameters could have taken a number of different values, finding that they are set within an extremely narrow range predicted by Eq.~\ref{eq.large} is even stronger evidence that your world is biased.  For example, the value of the cosmological constant in our world is more finely-tuned to our existence than it needs to be and is in close proximity to the value that maximizes Eq.~\ref{eq.large}\cite{Weinberg1987,Vilenkin1995,Weinberg2007,Pogosian2007}.  

\textbf{(3)} There will be circumstances where a natural process in your world has sometimes influenced the probability of your existence and othertimes where it has not.  Bias can be established by comparing the statistical properties of the process across these two cases.  For example, many of the properties of earth that make it especially conducive for life are statistical outliers when compared to other planets\cite{Ward2000, Waltham2014}.  If a small universe is unlikely to contain a planet as conducive for life as earth, then earth's existence is evidence that we live in a biased world.  Other potential examples include the following: evolutionary processes that have lead to our existence may exhibit different statistical properties than evolutionary processes that have not, and catastrophic risks (such as asteriod impacts) might occur much less frequently than baseline values determined from instances of these processes independent of our existence\cite{Cirkovic2010}.

\textbf{(4)} After you exist, there is no reason for processes in your world to remain biased towards your existence\cite{Tegmark2005}.  Therefore, in a large universe, a natural process that influences the probability of your existence will exhibit an abrupt change in its statistical properties after you exist.  In general, the process will appear conducive for your existence before you exist and will not after you exist.  Bias is established by comparing the properties of the process before and after you exist.  Boltzmann's fluctuation hypothesis for our world and the resulting arrow of time in thermodynamics is an example, albeit an incorrect one, of this statistical break\cite{Boltzmann1897}. (Boltzmann's hypothesis is incorrect because, as pointed out by Eddington\cite{Eddington1931}, the statistical break should occur at the moment of our existence, i.e., as part of a localized ``Boltzmann brain'' fluctuation, rather than at a much earlier moment in time.)  

It is this fourth method for establishing bias that I focus on below.

\section{Population Dynamics}

Assume that a solitary civilization of observers exists in all possible worlds.  The number of observers in the first generation of the civilization is $N_0$, and the size of later generations increases or decreases in an unbiased random way that is fully accounted for across possible worlds.  Generations are indexed by $t$, where $t={0,1,2,\dots,T}$.  The size of each generation, $N_t$, follows a random growth process,  
\begin{equation}
N_{t+1} = \Lambda_t N_t,
\end{equation}
where the growth factor, $\Lambda_t$, is a random variable that is equally likely to be $\lambda$ or $1/\lambda$.  Note that the continuous growth rate is zero on average,
\begin{equation}
\left\langle r_{t,t+\Delta t} \right\rangle = \left\langle \log{\left(\frac{N_{t+\Delta t}}{N_t}\right)} \middle/ \Delta t \right\rangle = 0.
\end{equation}

\begin{figure*}
\centering
\includegraphics[width=6.5in]{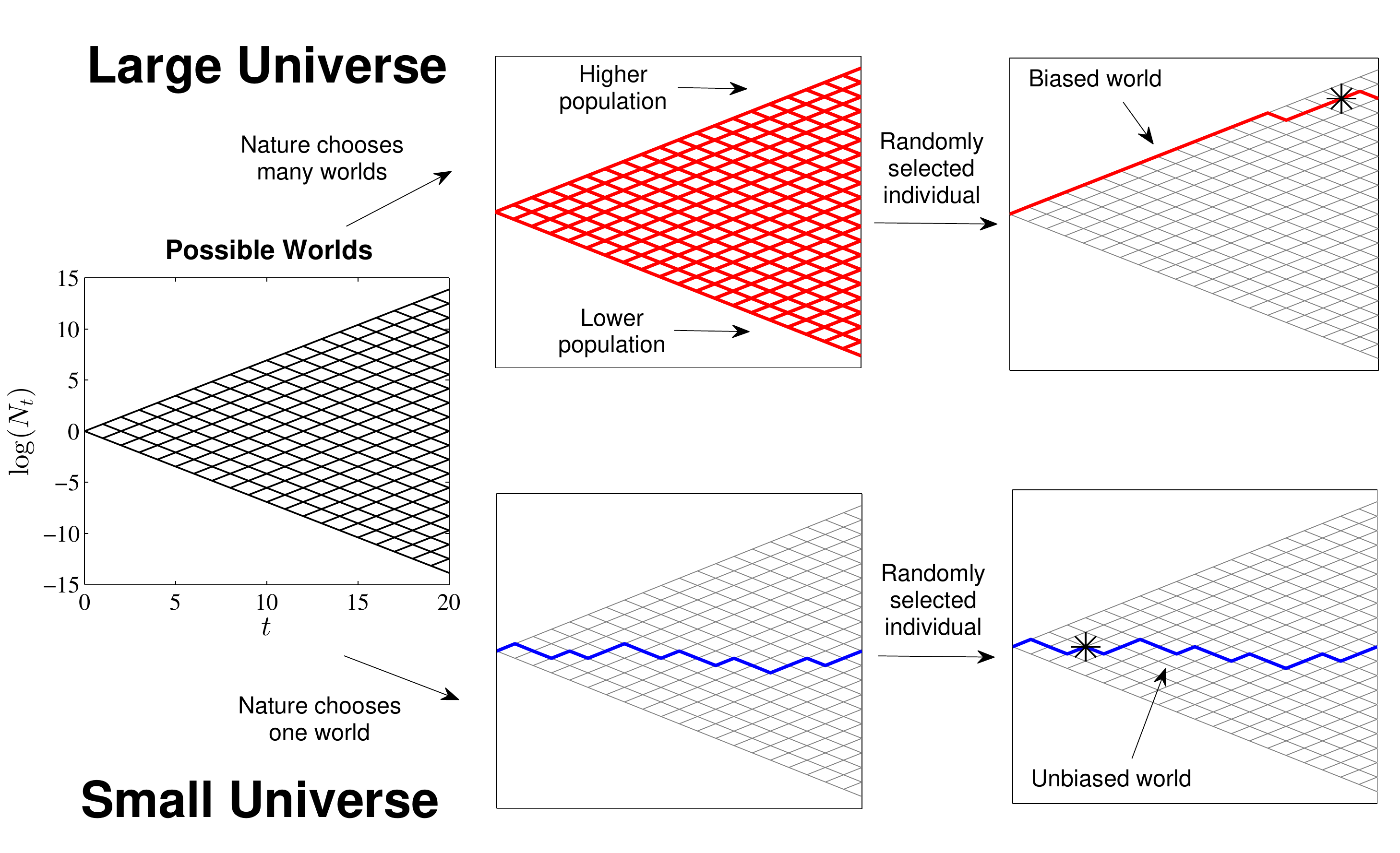}
\caption{Diagram illustrating how an observer in a large universe typically will exist in a world biased towards their existence, whereas an observer in a small universe typically will not.  The possible worlds represent different paths of population growth for a solitary civilization as modelled in the text.  Parameters for the model are $N_0=1$, $T$=20, and $\lambda=2$.}
\end{figure*}

\subsection{Example}

As an example, assume that $N_0=1$ million, $\lambda=2$, and $T=2$.  There are 4 paths for the population to follow and therefore 4 types of possible worlds to consider, $\{W_1,W_2,W_3,W_4\}$, which correspond to the respective population paths, $\{ \uparrow\uparrow, \uparrow\downarrow, \downarrow\uparrow, \downarrow\downarrow \}$.  For $\lambda$ and $1/\lambda$ to be equiprobable, each type of world must be equiprobable, $\mathcal{P}(\omega\in W_i) = 1/4$ for $i=1,\dots,4$.  Therefore, the probability that an observer exists in the different worlds in a small and large universe are (using Eqs.~\ref{eq.small} and \ref{eq.large}),
\begin{equation}
P_S(W_i)=\frac{1}{4} = 0.25 \ \ \mbox{ for } \ \ i=1,\dots,4,
\end{equation}
and,
\begin{eqnarray}
P_L(W_1) & = 7/15.25 & \approx 0.459,\\
P_L(W_2) & = 4/15.25 & \approx 0.262,\\
P_L(W_3) & = 2.5/15.25 & \approx 0.164,\\
P_L(W_4) & = 1.75/15.25 & \approx 0.115.
\end{eqnarray}

We can also calculate the probability that an observer exists at locations with specific properties.  For example, let the property $x$ represent time periods that correspond to each generation, $x\in\{0,1,2\}$.  The expected time-period in which an observer lives is,
\begin{eqnarray}
\left\langle x \right\rangle_S & = & 1,\\
\left\langle x \right\rangle_L & \approx & 1.15,
\end{eqnarray}
where $\left\langle \cdot \right\rangle_S$ and $\left\langle \cdot \right\rangle_L$ are averages taken over $P_S$ and $P_L$ respectively.  Notice that in the above model, an observer in a large universe is more likely to exist in worlds that produce more observers (Eqs.~11-14) and is more likely to exist in a later generation (Eq.~16).

In Fig.~1, I plot the set of possible population paths for $N_0=1$, $T$=20, and $\lambda=2$, and I demonstrate how an observer will typically find themselves in a biased world if the universe is large.  In a large universe, many worlds (and many civilizations) exist, and a randomly selected individual is much more likely to be drawn from a world with a large population, i.e., in a world biased towards their existence.  Using the same parameters as in Fig.~1, I show in Fig.~2 the probability that a randomly selected individual exists at different locations in a small and large universe, $P_S(x,y)$ and $P_L(x,y)$, where $x$ is the generation of an observer and $y$ is the logarithm of the size of an observer's generation.  The expected generation size through time (black line) and the average overall location of an observer (asterisk) in both a small and large universe are alse shown.   

Notice that even in a small universe, the location of an observer is biased.  The bias exists only within each world, towards locations that contain more observers (note that the distribution is slightly skewed upwards towards larger $N_t$ in Fig.~2(a)).  Because the bias in a small universe is only within each world, it is not as strong as the bias that exists in a large universe.

An observer who exists in a large universe can establish that their world is biased by comparing estimates of the continuous growth rate of the population measured before, $r_{0,x}$, and after, $r_{x,T}$, they exist.  In expectation, such an observer will measure the following,
\begin{eqnarray}
\left\langle r_{0,x} \right\rangle_L & \approx & 0.416,\\
\left\langle r_{x,T} \right\rangle_L & = & 0.
\end{eqnarray}
Notice that the growth rate of the population is biased before a typical observer exists, but reverts to its baseline value after they exist.  A similar effect exists in a small universe, although the nature of the bias is different (the bias is first positive and then negative), and is less pronounced: $\left\langle r_{0,x} \right\rangle_S \approx 0.115$ and $\left\langle r_{0,x} \right\rangle_S \approx -0.115$

\begin{figure*}
\centering
\includegraphics[width=3.25in]{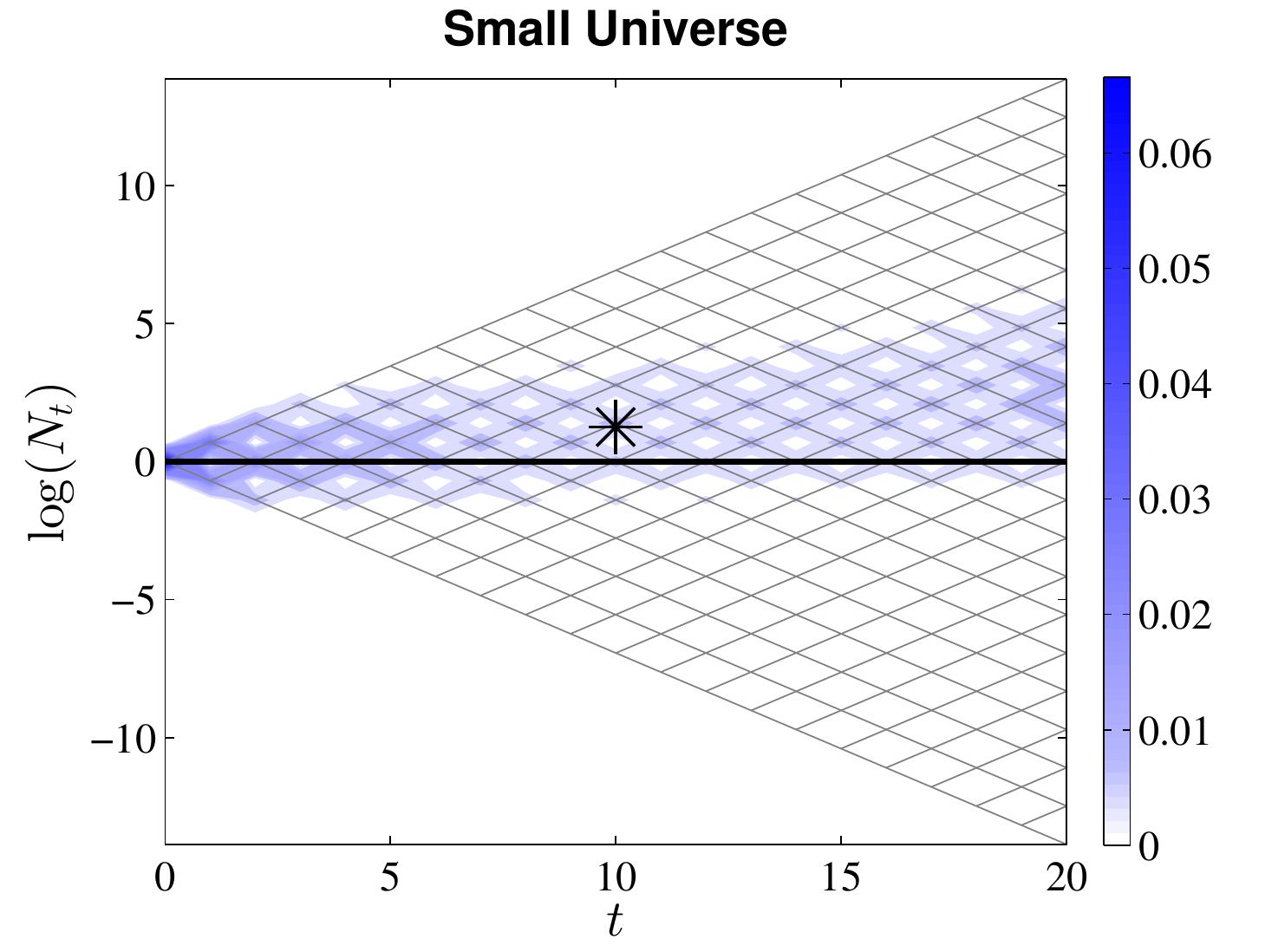}
\includegraphics[width=3.25in]{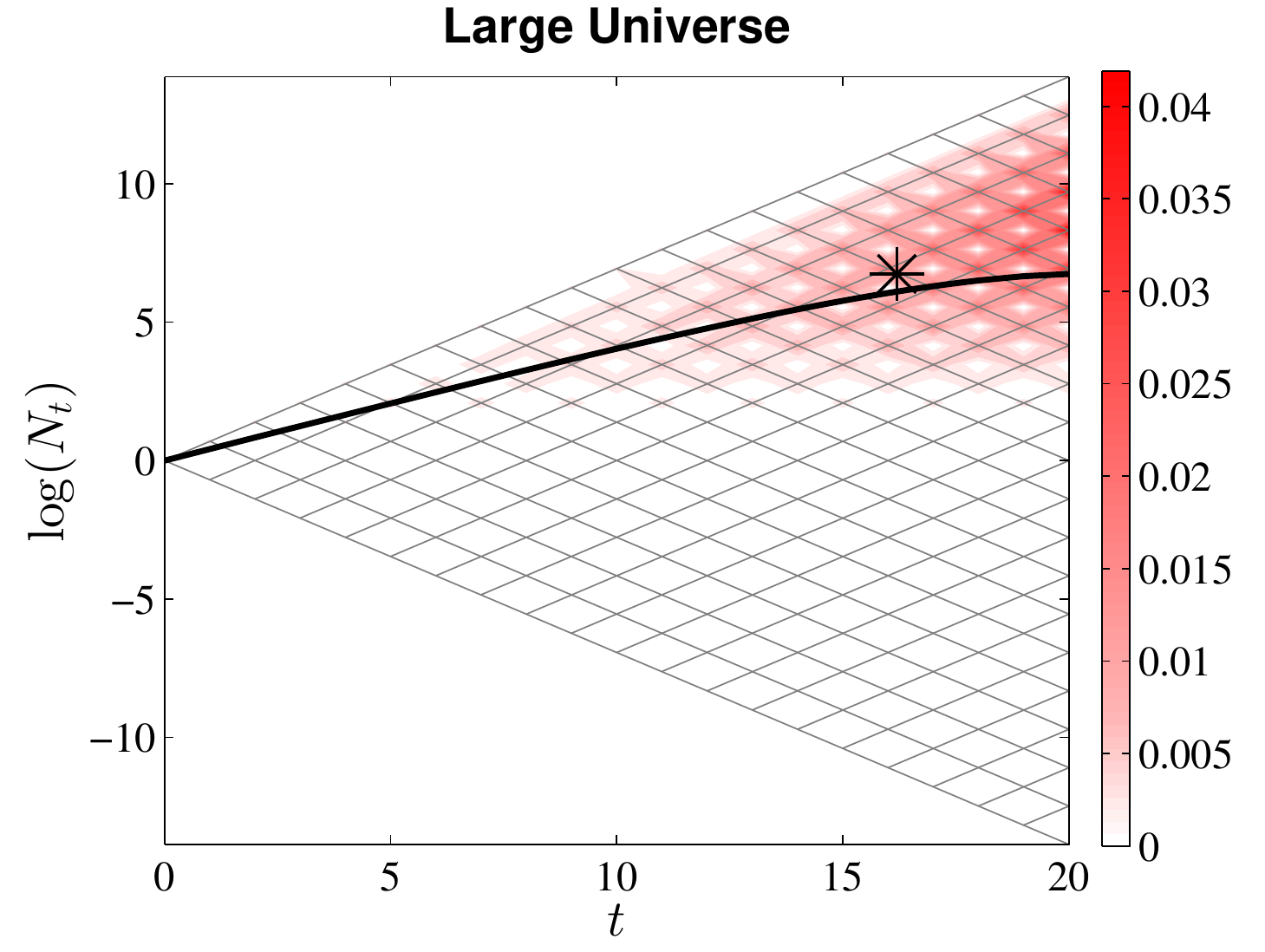}
\caption{The probability of existing at location $x=t$ and $y=Log(N_t))$ in (a) a small universe and (b) a large universe assuming the population model described in the text with parameters $N_0=1$, $T$=20, and $\lambda=2$.  The expected generation size through time (black line) and the average overall location of an observer (asterisk) are alse shown.  In the figures, $\left\langle x \right\rangle_S = 10$, $\left\langle y \right\rangle_S = 1.25$, $\left\langle x \right\rangle_L = 16.2$, $\left\langle y \right\rangle_L = 6.74$.}
\end{figure*}

\subsection{Our Population}

Although the above model of population dynamics is quite crude, we can attempt to fit it to our own population for illustration.  In Fig.~3(a) I plot estimates of the logarithm of human birth rate (total births per year) at 40,000 BC, 6500 BC, 1 AD, and 2010 AD.  In addition, I show the expected generation size through time (black line) and the standard deviation around this expectation (dotted black lines) for the population model with parameters $\lambda=3.3$, $n_0=1$, and $T=20$.  To enforce correspondence between the model and the data, the parameters of the model were chosen so that the expected generation size of an observer in the model is equal to our current birth rate (total births per year), and the horizontal axis is normalized so that the year 2010 corresponds to the expected generation of an observer in the model.

In Fig.~3(b) and (c) I show estimates and projections of birth rates from 1650 to 2100 and fertility rates from 1950 through 2100 based on United Nations data.  Birth rates increased rather quickly during the industrial revolution, but this growth abruptly stopped over the last 50 years.

\begin{figure*}
\centering
\includegraphics[width=3.1in]{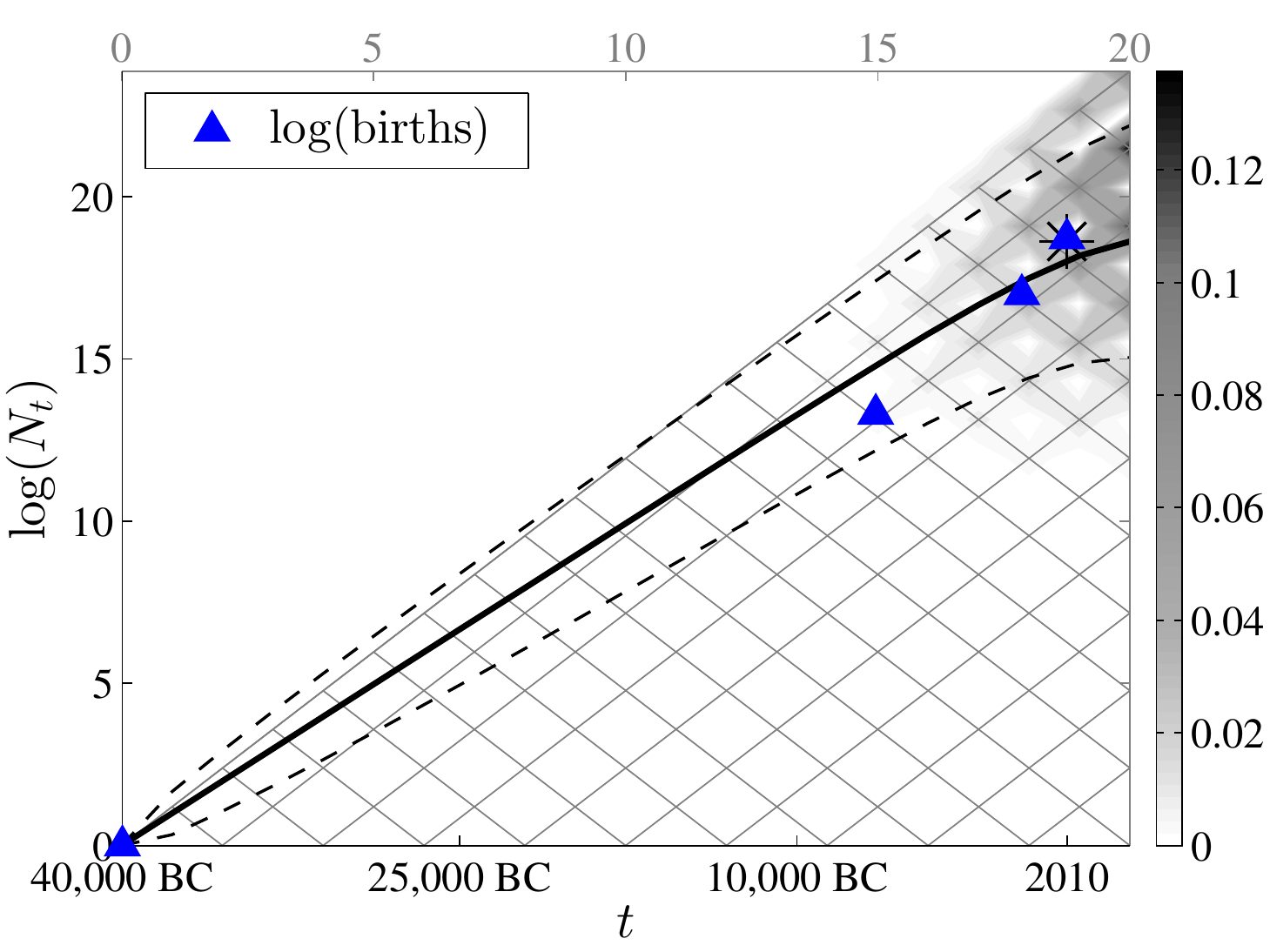}
\includegraphics[width=3.1in]{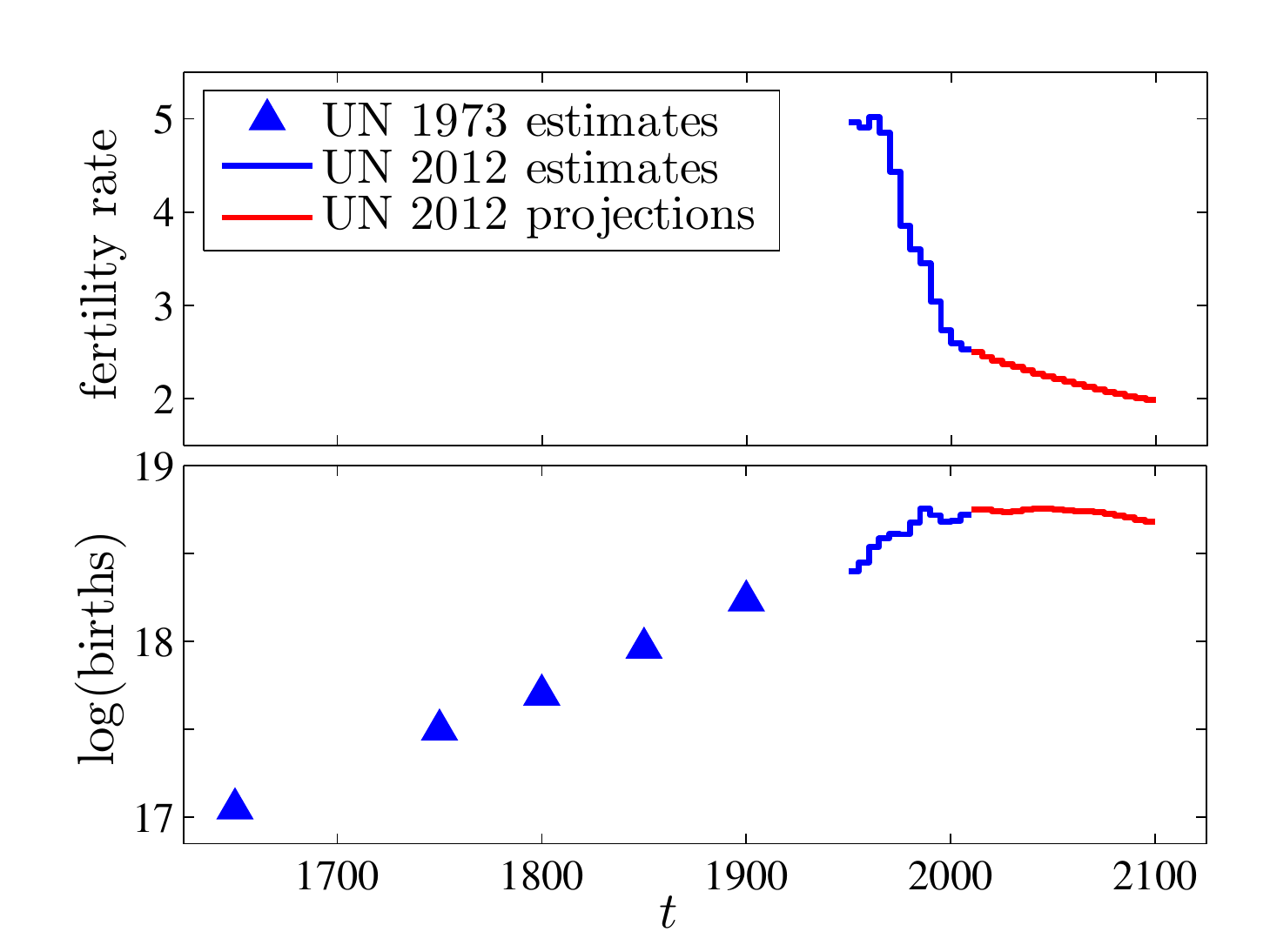}
\caption{(a) Human births per year (blue triangles) compared to the population model described in the text with parameters $N_0=1$, $T$=20, and $\lambda=3.3$.  The expected generation size through time (black line), the standard deviation around this expectation (black dashed line), and the average overall location of an observer (asterisk) in the model are shown.  The horizontal axis measures time and is in units of years (bottom) and periods of the model (top).  (b) United Nations estimates and projections of global fertility from 1950 through 2100.  (c) United Nations estimates and projections of births per year from 1650 through 2100.}
\end{figure*}

Notice that our circumstance is consistent with that of a typical observer whose population is naturally steady through time and who lives in a large universe.  Such an observer would measure positive population growth in the past which is removed at the time of their birth. 

Of course, it is difficult to consider this consistency as strong evidence of many worlds.  It could be that our population growth and subsequent halt is a typical path for a civilization like ours, or alternatively, that an extremely large number of civilizations exist in a small universe (which would produce the same effect).   On the other hand, the correspondence between the model and our circumstance should not be dismissed.  We do not have evidence that other civilizations in our world exist and it is not at all clear that our population dynamics have been typical (for instance, consider that HIV could have spread to humans at any point in our history).

\section{Conclusions}
Anthropic reasoning has been used in cosmology for many years\cite{Eddington1931}, long before it was codified by Carter in 1974\cite{Carter1974}.  It helps us decide on theories: theories that are incompatible with our existence cannot be valid and we should exclude them.  In addition, we should discount theories that do not make our circumstance typical, and the ``complete anthropic principle'' presented here now makes this clear.

Carter was quick to point out that the anthropic principle applies beyond cosmology and constrains theories in biology as well\cite{Carter1983}.  Likewise, the complete anthropic principle has implications outside of cosmology.  If the universe is large, then the typical circumstance of an observer will likely be atypical according to the laws of nature, and all disciplines must be aware of and correct for this bias.

The atypical situation of a typical observer also has important implications for how science operates.  Usually when considering a theory, scientists start from initial conditions and move forward along typically produced paths.  If there are many worlds, then this paradigm is incorrect.  Instead of finding ourselves in a typical world according to the laws of nature (Eq.~2), each one of us should find ourselves living in a world that made our existence extremely likely (Eq.~3).  Scientific explanation, therefore must start with each of us separately, working outward and searching for those worlds that typically produce us rather than those worlds that are typically produced.

\begin{acknowledgments}
I thank Ken Olum, Don Page, Rainer Plaga, and Alex Vilenkin for very helpful comments and suggestions.
\end{acknowledgments}

\bibliography{AreThereManyWorlds}

\begin{thebibliography}{43}
\expandafter\ifx\csname natexlab\endcsname\relax\def\natexlab#1{#1}\fi
\expandafter\ifx\csname bibnamefont\endcsname\relax
  \def\bibnamefont#1{#1}\fi
\expandafter\ifx\csname bibfnamefont\endcsname\relax
  \def\bibfnamefont#1{#1}\fi
\expandafter\ifx\csname citenamefont\endcsname\relax
  \def\citenamefont#1{#1}\fi
\expandafter\ifx\csname url\endcsname\relax
  \def\url#1{\texttt{#1}}\fi
\expandafter\ifx\csname urlprefix\endcsname\relax\def\urlprefix{URL }\fi
\providecommand{\bibinfo}[2]{#2}
\providecommand{\eprint}[2][]{\url{#2}}

\bibitem[{\citenamefont{Everett}(1957)}]{Everett1957}
\bibinfo{author}{\bibfnamefont{H.}~\bibnamefont{Everett}},
  \bibinfo{journal}{Review of Modern Physics} \textbf{\bibinfo{volume}{29}},
  \bibinfo{pages}{454} (\bibinfo{year}{1957}).

\bibitem[{\citenamefont{DeWitt and Graham}(1973)}]{DeWitt1973}
\bibinfo{author}{\bibfnamefont{B.~S.} \bibnamefont{DeWitt}} \bibnamefont{and}
  \bibinfo{author}{\bibfnamefont{N.}~\bibnamefont{Graham}},
  \emph{\bibinfo{title}{The Many-Worlds Interpretation of Quantum Mechanics}}
  (\bibinfo{publisher}{Princeton University Press},
  \bibinfo{address}{Princeton, NJ}, \bibinfo{year}{1973}).

\bibitem[{\citenamefont{Guth}(1997)}]{Guth1997}
\bibinfo{author}{\bibfnamefont{A.~H.} \bibnamefont{Guth}},
  \emph{\bibinfo{title}{The Inflationary Universe: The Quest for a New Theory
  of Cosmic Origins}} (\bibinfo{publisher}{Perseus Books},
  \bibinfo{year}{1997}).

\bibitem[{\citenamefont{Garriga and Vilenkin}(2001)}]{Garriga2001}
\bibinfo{author}{\bibfnamefont{J.}~\bibnamefont{Garriga}} \bibnamefont{and}
  \bibinfo{author}{\bibfnamefont{A.}~\bibnamefont{Vilenkin}},
  \bibinfo{journal}{Physical Review D} \textbf{\bibinfo{volume}{64}},
  \bibinfo{pages}{043511} (\bibinfo{year}{2001}).

\bibitem[{\citenamefont{Tegmark}(2003)}]{Tegmark2003}
\bibinfo{author}{\bibfnamefont{M.}~\bibnamefont{Tegmark}},
  \bibinfo{journal}{Scientific American} \textbf{\bibinfo{volume}{288}},
  \bibinfo{pages}{40} (\bibinfo{year}{2003}).

\bibitem[{\citenamefont{Vilenkin}(2006)}]{Vilenkin2006}
\bibinfo{author}{\bibfnamefont{A.}~\bibnamefont{Vilenkin}},
  \emph{\bibinfo{title}{Many Worlds in One: The Search for Other Universes}}
  (\bibinfo{publisher}{Hill and Wang}, \bibinfo{address}{New York},
  \bibinfo{year}{2006}).

\bibitem[{\citenamefont{Carter}(1974)}]{Carter1974}
\bibinfo{author}{\bibfnamefont{B.}~\bibnamefont{Carter}}, in
  \emph{\bibinfo{booktitle}{Confrontation of Cosmological Theories with
  Observational Data}}, edited by \bibinfo{editor}{\bibfnamefont{M.~S.}
  \bibnamefont{Longair}} (\bibinfo{publisher}{IAU}, \bibinfo{year}{1974}).

\bibitem[{\citenamefont{Carter}(1983)}]{Carter1983}
\bibinfo{author}{\bibfnamefont{B.}~\bibnamefont{Carter}},
  \bibinfo{journal}{Philosophical Transactions of the Royal Society A}
  \textbf{\bibinfo{volume}{310}}, \bibinfo{pages}{347} (\bibinfo{year}{1983}).

\bibitem[{\citenamefont{Dicke}(1961)}]{Dicke1961}
\bibinfo{author}{\bibfnamefont{R.~H.} \bibnamefont{Dicke}},
  \bibinfo{journal}{Nature} \textbf{\bibinfo{volume}{192}},
  \bibinfo{pages}{440} (\bibinfo{year}{1961}).

\bibitem[{\citenamefont{Eddington}(1931)}]{Eddington1931}
\bibinfo{author}{\bibfnamefont{A.~S.} \bibnamefont{Eddington}},
  \bibinfo{journal}{Nature} \textbf{\bibinfo{volume}{127}},
  \bibinfo{pages}{447} (\bibinfo{year}{1931}).

\bibitem[{\citenamefont{Barrow and Tipler}(1986)}]{Barrow1986}
\bibinfo{author}{\bibfnamefont{J.~D.} \bibnamefont{Barrow}} \bibnamefont{and}
  \bibinfo{author}{\bibfnamefont{F.~J.} \bibnamefont{Tipler}},
  \emph{\bibinfo{title}{The Anthropic Cosmological Principle}}
  (\bibinfo{publisher}{Oxford University Press}, \bibinfo{address}{New York},
  \bibinfo{year}{1986}).

\bibitem[{\citenamefont{Vilenkin}(1995)}]{Vilenkin1995}
\bibinfo{author}{\bibfnamefont{A.}~\bibnamefont{Vilenkin}},
  \bibinfo{journal}{Physical Review Letters} \textbf{\bibinfo{volume}{74}},
  \bibinfo{pages}{846} (\bibinfo{year}{1995}).

\bibitem[{\citenamefont{Leslie}(1990)}]{Leslie1990}
\bibinfo{author}{\bibfnamefont{J.}~\bibnamefont{Leslie}},
  \bibinfo{journal}{Interchange} \textbf{\bibinfo{volume}{21}},
  \bibinfo{pages}{49} (\bibinfo{year}{1990}).

\bibitem[{\citenamefont{Olum}(2002)}]{Olum2000}
\bibinfo{author}{\bibfnamefont{K.~D.} \bibnamefont{Olum}},
  \bibinfo{journal}{Philosophical Quarterly} \textbf{\bibinfo{volume}{52}},
  \bibinfo{pages}{164} (\bibinfo{year}{2002}).

\bibitem[{\citenamefont{Bostrom}(2002)}]{Bostrom2002}
\bibinfo{author}{\bibfnamefont{N.}~\bibnamefont{Bostrom}},
  \emph{\bibinfo{title}{Anthropic Bias: Observation Selection Effects}}
  (\bibinfo{publisher}{Routledge}, \bibinfo{address}{New York},
  \bibinfo{year}{2002}).

\bibitem[{\citenamefont{Bostrom and \'{C}irkovi\'{c}}(2003)}]{Bostrom2003}
\bibinfo{author}{\bibfnamefont{N.}~\bibnamefont{Bostrom}} \bibnamefont{and}
  \bibinfo{author}{\bibfnamefont{M.~M.} \bibnamefont{\'{C}irkovi\'{c}}},
  \bibinfo{journal}{Philosophical Quarterly} \textbf{\bibinfo{volume}{53}},
  \bibinfo{pages}{83} (\bibinfo{year}{2003}).

\bibitem[{\citenamefont{Albrecht and Sorbo}(2004)}]{Albrecht2004}
\bibinfo{author}{\bibfnamefont{A.}~\bibnamefont{Albrecht}} \bibnamefont{and}
  \bibinfo{author}{\bibfnamefont{L.}~\bibnamefont{Sorbo}},
  \bibinfo{journal}{Physical Review D} \textbf{\bibinfo{volume}{70}},
  \bibinfo{pages}{063528} (\bibinfo{year}{2004}).

\bibitem[{\citenamefont{Tegmark and Bostrom}(2005)}]{Tegmark2005}
\bibinfo{author}{\bibfnamefont{M.}~\bibnamefont{Tegmark}} \bibnamefont{and}
  \bibinfo{author}{\bibfnamefont{N.}~\bibnamefont{Bostrom}},
  \bibinfo{journal}{Nature} \textbf{\bibinfo{volume}{438}},
  \bibinfo{pages}{754} (\bibinfo{year}{2005}).

\bibitem[{\citenamefont{Neal}(2006)}]{Neal2006}
\bibinfo{author}{\bibfnamefont{R.~M.} \bibnamefont{Neal}}
  (\bibinfo{year}{2006}), \bibinfo{note}{e-print arXiv:math/0608592}.

\bibitem[{\citenamefont{Page}(2007)}]{Page2007}
\bibinfo{author}{\bibfnamefont{D.~N.} \bibnamefont{Page}}
  (\bibinfo{year}{2007}), \bibinfo{note}{e-print arXiv:0707.4169}.

\bibitem[{\citenamefont{Weinberg}(2007)}]{Weinberg2007}
\bibinfo{author}{\bibfnamefont{S.}~\bibnamefont{Weinberg}}, in
  \emph{\bibinfo{booktitle}{Universe or Multiverse?}}, edited by
  \bibinfo{editor}{\bibfnamefont{B.}~\bibnamefont{Carr}}
  (\bibinfo{publisher}{Cambridge University Press},
  \bibinfo{address}{Cambridge}, \bibinfo{year}{2007}).

\bibitem[{\citenamefont{Pogosian and Vilenkin}(2007)}]{Pogosian2007}
\bibinfo{author}{\bibfnamefont{L.}~\bibnamefont{Pogosian}} \bibnamefont{and}
  \bibinfo{author}{\bibfnamefont{A.}~\bibnamefont{Vilenkin}},
  \bibinfo{journal}{JCAP} \textbf{\bibinfo{volume}{01}}, \bibinfo{pages}{025}
  (\bibinfo{year}{2007}).

\bibitem[{\citenamefont{Page}(2008)}]{Page2008}
\bibinfo{author}{\bibfnamefont{D.~N.} \bibnamefont{Page}},
  \bibinfo{journal}{Physics Letters B} \textbf{\bibinfo{volume}{669}},
  \bibinfo{pages}{197} (\bibinfo{year}{2008}).

\bibitem[{\citenamefont{Garriga and Vilenkin}(2008)}]{Garriga2008}
\bibinfo{author}{\bibfnamefont{J.}~\bibnamefont{Garriga}} \bibnamefont{and}
  \bibinfo{author}{\bibfnamefont{A.}~\bibnamefont{Vilenkin}},
  \bibinfo{journal}{Physical Review D} \textbf{\bibinfo{volume}{77}},
  \bibinfo{pages}{043526} (\bibinfo{year}{2008}).

\bibitem[{\citenamefont{Bousso et~al.}(2008)\citenamefont{Bousso, Freivogel,
  and Yang}}]{Bousso2008}
\bibinfo{author}{\bibfnamefont{R.}~\bibnamefont{Bousso}},
  \bibinfo{author}{\bibfnamefont{B.}~\bibnamefont{Freivogel}},
  \bibnamefont{and} \bibinfo{author}{\bibfnamefont{I.-S.} \bibnamefont{Yang}},
  \bibinfo{journal}{Physical Review D} \textbf{\bibinfo{volume}{77}},
  \bibinfo{pages}{103514} (\bibinfo{year}{2008}).

\bibitem[{\citenamefont{\'{C}irkovi\'{c}
  et~al.}(2010)\citenamefont{\'{C}irkovi\'{c}, Sandberg, and
  Bostrom}}]{Cirkovic2010}
\bibinfo{author}{\bibfnamefont{M.~M.} \bibnamefont{\'{C}irkovi\'{c}}},
  \bibinfo{author}{\bibfnamefont{A.}~\bibnamefont{Sandberg}}, \bibnamefont{and}
  \bibinfo{author}{\bibfnamefont{N.}~\bibnamefont{Bostrom}},
  \bibinfo{journal}{Risk Analysis} \textbf{\bibinfo{volume}{30}},
  \bibinfo{pages}{1495} (\bibinfo{year}{2010}).

\bibitem[{\citenamefont{Waltham}(2011)}]{Waltham2011}
\bibinfo{author}{\bibfnamefont{D.}~\bibnamefont{Waltham}},
  \bibinfo{journal}{Icarus} \textbf{\bibinfo{volume}{215}},
  \bibinfo{pages}{518} (\bibinfo{year}{2011}).

\bibitem[{\citenamefont{Gerig}(2012)}]{Gerig2012}
\bibinfo{author}{\bibfnamefont{A.}~\bibnamefont{Gerig}} (\bibinfo{year}{2012}),
  \bibinfo{note}{e-print arXiv:1209.6251}.

\bibitem[{\citenamefont{Gerig et~al.}(2013)\citenamefont{Gerig, Olum, and
  Vilenkin}}]{Gerig2013}
\bibinfo{author}{\bibfnamefont{A.}~\bibnamefont{Gerig}},
  \bibinfo{author}{\bibfnamefont{K.~D.} \bibnamefont{Olum}}, \bibnamefont{and}
  \bibinfo{author}{\bibfnamefont{A.}~\bibnamefont{Vilenkin}},
  \bibinfo{journal}{JCAP} \textbf{\bibinfo{volume}{05}}, \bibinfo{pages}{013}
  (\bibinfo{year}{2013}).

\bibitem[{\citenamefont{Hartle and Srednicki}(2007)}]{Hartle2007}
\bibinfo{author}{\bibfnamefont{J.~B.} \bibnamefont{Hartle}} \bibnamefont{and}
  \bibinfo{author}{\bibfnamefont{M.}~\bibnamefont{Srednicki}},
  \bibinfo{journal}{Physical Review D} \textbf{\bibinfo{volume}{75}},
  \bibinfo{pages}{123523} (\bibinfo{year}{2007}).

\bibitem[{\citenamefont{Srednicki and Hartle}(2010)}]{Srednicki2010}
\bibinfo{author}{\bibfnamefont{M.}~\bibnamefont{Srednicki}} \bibnamefont{and}
  \bibinfo{author}{\bibfnamefont{J.}~\bibnamefont{Hartle}},
  \bibinfo{journal}{Physical Review D} \textbf{\bibinfo{volume}{81}},
  \bibinfo{pages}{123524} (\bibinfo{year}{2010}).

\bibitem[{\citenamefont{Weinberg}(1987)}]{Weinberg1987}
\bibinfo{author}{\bibfnamefont{S.}~\bibnamefont{Weinberg}},
  \bibinfo{journal}{Physical Review Letters} \textbf{\bibinfo{volume}{59}},
  \bibinfo{pages}{2607} (\bibinfo{year}{1987}).

\bibitem[{\citenamefont{Leslie}(1988)}]{Leslie1988}
\bibinfo{author}{\bibfnamefont{J.}~\bibnamefont{Leslie}},
  \bibinfo{journal}{Mind} \textbf{\bibinfo{volume}{97}}, \bibinfo{pages}{269}
  (\bibinfo{year}{1988}).

\bibitem[{\citenamefont{van Inwagen}(1993)}]{vanInwagen1993}
\bibinfo{author}{\bibfnamefont{P.}~\bibnamefont{van Inwagen}},
  \emph{\bibinfo{title}{Metaphysics}} (\bibinfo{publisher}{Westview Press},
  \bibinfo{address}{Boulder}, \bibinfo{year}{1993}).

\bibitem[{\citenamefont{Bradley}(2009)}]{Bradley2009}
\bibinfo{author}{\bibfnamefont{D.}~\bibnamefont{Bradley}},
  \bibinfo{journal}{American Philosophical Quarterly}
  \textbf{\bibinfo{volume}{46}}, \bibinfo{pages}{61} (\bibinfo{year}{2009}).

\bibitem[{\citenamefont{Carr and Rees}(1979)}]{Carr1979}
\bibinfo{author}{\bibfnamefont{B.~J.} \bibnamefont{Carr}} \bibnamefont{and}
  \bibinfo{author}{\bibfnamefont{M.~J.} \bibnamefont{Rees}},
  \bibinfo{journal}{Nature} \textbf{\bibinfo{volume}{278}},
  \bibinfo{pages}{605} (\bibinfo{year}{1979}).

\bibitem[{\citenamefont{Hogan}(2000)}]{Hogan2000}
\bibinfo{author}{\bibfnamefont{C.~J.} \bibnamefont{Hogan}},
  \bibinfo{journal}{Reviews of Modern Physics} \textbf{\bibinfo{volume}{72}},
  \bibinfo{pages}{1149} (\bibinfo{year}{2000}).

\bibitem[{\citenamefont{Page}(1999{\natexlab{a}})}]{Page1999}
\bibinfo{author}{\bibfnamefont{D.~N.} \bibnamefont{Page}},
  \bibinfo{journal}{AIP Conference Proceedings} \textbf{\bibinfo{volume}{493}},
  \bibinfo{pages}{225} (\bibinfo{year}{1999}{\natexlab{a}}).

\bibitem[{\citenamefont{Page}(1999{\natexlab{b}})}]{Page1999b}
\bibinfo{author}{\bibfnamefont{D.~N.} \bibnamefont{Page}}
  (\bibinfo{year}{1999}{\natexlab{b}}), \bibinfo{note}{e-print
  arXiv:quant-ph/9904004}.

\bibitem[{\citenamefont{Page}(2011)}]{Page2011}
\bibinfo{author}{\bibfnamefont{D.~N.} \bibnamefont{Page}}
  (\bibinfo{year}{2011}), \bibinfo{note}{e-print arXiv:1101.1083}.

\bibitem[{\citenamefont{Ward and Brownlee}(2000)}]{Ward2000}
\bibinfo{author}{\bibfnamefont{P.~D.} \bibnamefont{Ward}} \bibnamefont{and}
  \bibinfo{author}{\bibfnamefont{D.}~\bibnamefont{Brownlee}},
  \emph{\bibinfo{title}{Rare Earth: Why Complex Life is Uncommon in the
  Universe}} (\bibinfo{publisher}{Springer}, \bibinfo{address}{New York},
  \bibinfo{year}{2000}).

\bibitem[{\citenamefont{Waltham}(2014)}]{Waltham2014}
\bibinfo{author}{\bibfnamefont{D.}~\bibnamefont{Waltham}},
  \emph{\bibinfo{title}{Lucky Planet: Why Earth is Exceptional -- and What That
  Means for Life in the Universe}} (\bibinfo{publisher}{Basic Books},
  \bibinfo{address}{New York}, \bibinfo{year}{2014}).

\bibitem[{\citenamefont{Boltzmann}(1897)}]{Boltzmann1897}
\bibinfo{author}{\bibfnamefont{L.}~\bibnamefont{Boltzmann}},
  \bibinfo{journal}{Annalen der Physik} \textbf{\bibinfo{volume}{60}},
  \bibinfo{pages}{392} (\bibinfo{year}{1897}).

\end{thebibliography}

\end{document}